\documentclass[psfig,useAMS,usenatbib]{mn2e}
\usepackage{times}
\usepackage{epsfig}
\usepackage{amsmath}
\usepackage{natbib}
\usepackage{longtable}

\def\apj {ApJ}
\def\apjl {ApJL}
\def\apjs {ApJS}
\def\aj {AJ}
\def\aap {A\&A}
\def\mnras {MNRAS}
\def\pasp {PASP}
\def\zphot {z_{\rm phot}}
\def\zspec {z_{\rm spec}}
\title[Evolution of environment dependent galaxy properties in the SDSS]
  {Evolution of environment dependent galaxy properties in the Sloan Digital Sky Survey}
\author[O'Mill et al.]
  {Ana Laura O'Mill,$^{1}$\thanks{E-mail: aomill@oac.uncor.edu}
  Nelson Padilla,$^{2}$ Diego Garc\'\i a Lambas,$^{1,3}$\\
  $^1$Grupo de Investigaciones en Astronom\'\i a Te\'orica y 
Experimental, IATE, Observatorio Astron\'omico, Universidad Nacional de C\'ordoba,\\ 
Laprida 854, X5000BGR, C\'ordoba Argentina\\
  $^2$ 2 Departamento de Astronom\'\i a y Astrof\'\i sica, Pontificia Universidad Cat\'olica de Chile, 
V. Mackenna 4860, Santiago 22, Chile.\\
  $^3$Consejo de Investigaciones Cient\'\i ficas y T\'ecnicas (CONICET), Avenida Rivadavia 
1917, C1033AAJ, Buenos Aires, Argentina}
\date{Released 2006 Xxxxx XX}
\pagerange{\pageref{firstpage}--\pageref{lastpage}} \pubyear{2002}
\def\LaTeX{L\kern-.36em\raise.3ex\hbox{a}\kern-.15em
    T\kern-.1667em\lower.7ex\hbox{E}\kern-.125emX}

\begin{document}
\label{firstpage}
\maketitle
\begin{abstract}
We use photometric redshifts to analyse the effect of local environment 
on galaxy colours at redshifts $ z \la 0.63$ in the SDSS data release 6.  
We construct mock SDSS-DR6 catalogues using semi-analytic galaxies to
study possible systematic effects on the characterisation of environment and
colour statistics due to the uncertainty in the determination of redshifts.
We use the projected galaxy 
density derived from the distance to the nearest neighbours
with a suitable radial velocity threshold 
to take into account the uncertainties in the photometric redshift estimates.
Our findings indicate that the use of photometric redshifts greatly 
improves estimates of projected local galaxy density when 
galaxy spectra are not available.  We find a tight relationship between
spectroscopic and photometric derived densities, both in the SDSS-DR6 
data (up to $z=0.3$) and mock catalogues (up to $z=0.63$).

At $z=0$, faint galaxies show a clear increase of the red galaxy fraction as
the local density increases.  Bright galaxies, on the other hand, show a constant
red galaxy fraction.  We are able to track
the evolution of this fraction to $z=0.55$ for galaxies brighter than
$M_r=-21.5$ and find that the fraction of blue galaxies with respect to the
total population progressively becomes higher
as the redshift increases, at a rate of $15\%/$Gyr.  Also, 
at any given redshift, bright galaxies show a larger red population, 
indicating that the star-formation activity shifts towards
smaller objects as the redshift decreases.
\end{abstract}
\section{Introduction}
The study of galaxies in the field and in clusters has revealed
the existence of significant correlations between several galaxy properties 
and their environment. 
In a  pioneering work by 
\citet{dress80}, it was shown for the first time that
Galaxy morphologies depend on the local galaxy density.
The decrease in the SFR of galaxies 
in dense environments is a universal phenomenon over a wide range of densities.  For instance,
\citet{gomez03} found that the star formation rate (SFR) of galaxies is 
strongly correlated with the local galaxy density;
\citet{balogh04} characterise the environment using projected and 
three-dimensional densities concluding that the present-day correlation 
between star formation rate and environment is a result of short-timescale 
mechanisms that take place preferentially at high redshift, such as 
starbursts induced by galaxy-galaxy interactions (see also \citealt{baldry04}). 
However, these studies 
concern the nearby universe where spectroscopic redshifts allow for estimates 
of the local density of galaxies. 
At higher redshifts, spectroscopy is not available 
for large galaxy samples, and in consequence, the relation between galaxy properties 
and environment is more poorly understood. An alternative solution to this problem 
is the use of multi-band photometry to constrain galaxy redshifts.
These techniques have been extensively studied (e.g., 
\citealt{koo85,con95,gh96,ben00,bolz00,csabai03,annz}), and
have proved to be efficient in estimating redshifts 
for large numbers of galaxies, which opens the possibility of obtaining their intrinsic properties
for statistical studies.

One exception to this necessity is that of the VIMOS VLT Deep Survey, for which Cucciati et al. (2006)
make use of the available spectroscopic redshifts to estimate local densities; however,
their sample is small containing only 6582 galaxies out to $z=1.5$.  Their results show that 
massive galaxies shift towards redder values for lower redshifts, where only faint galaxies
show a blue galaxy population.
On the other hand, the use of photometric redshifts to study the evolution of galaxy colours 
at redshifts higher than $z=0.3$, has only been attempted a few times, in part due
to the difficulty in ensuring an accurate statistical measurement of local density.
For instance, \citet{roser} study the colour distribution of galaxies in the CFHTLS-Deep
Field Survey
\footnote{http:$//$www.ast.obs-mip.fr/users/roser/CFHTLS\_T0003} 
for redshifts in the range $0<z<1.3$,
using a well calibrated photometric redshift estimator, New-HyperZ\footnote{
http:$//$www.ast.obs-mip.fr/users/roser/hyperz
}, to find results consistent with those from the VIMOS VLT Deep Survey, indicating
the validity of the photometric redshift approach.
De Lucia et al.  (2007), find that the population of cluster
galaxies in the Las Campanas Distant Cluster Survey (Gonzalez et al., 2001) shows a clear deficit
of blue galaxies in low redshift galaxies with respect to clusters at $z\sim0.8$.
The variation of galaxy colours as a function of redshift provides important information on
the stellar formation history of galaxies; De Lucia et al. fit their results with models where the star
formation is suppressed early on in hostile cluster environments.

In this paper we analyse galaxy properties and their relation to the environment
by means of photometric data taken from the Sloan Digital Sky Survey, Data Release 6 (SDSS-DR6,
\citealt{sdss}).  We use galaxy photometric 
redshifts, $\zphot$, from the SDSS database, to compute local densities, intrinsic
luminosities and colours.  The main advantage of our study is the comparatively large
number of galaxies in the SDSS-DR6, at redshifts $0.1<z<0.63$, a range where deep spectroscopic surveys
only offer a limited number of objects, and wide spectroscopic surveys are badly affected by
selection biases.

\begin{figure*}
\begin{picture}(450,240)
\put(0,0){\psfig{file=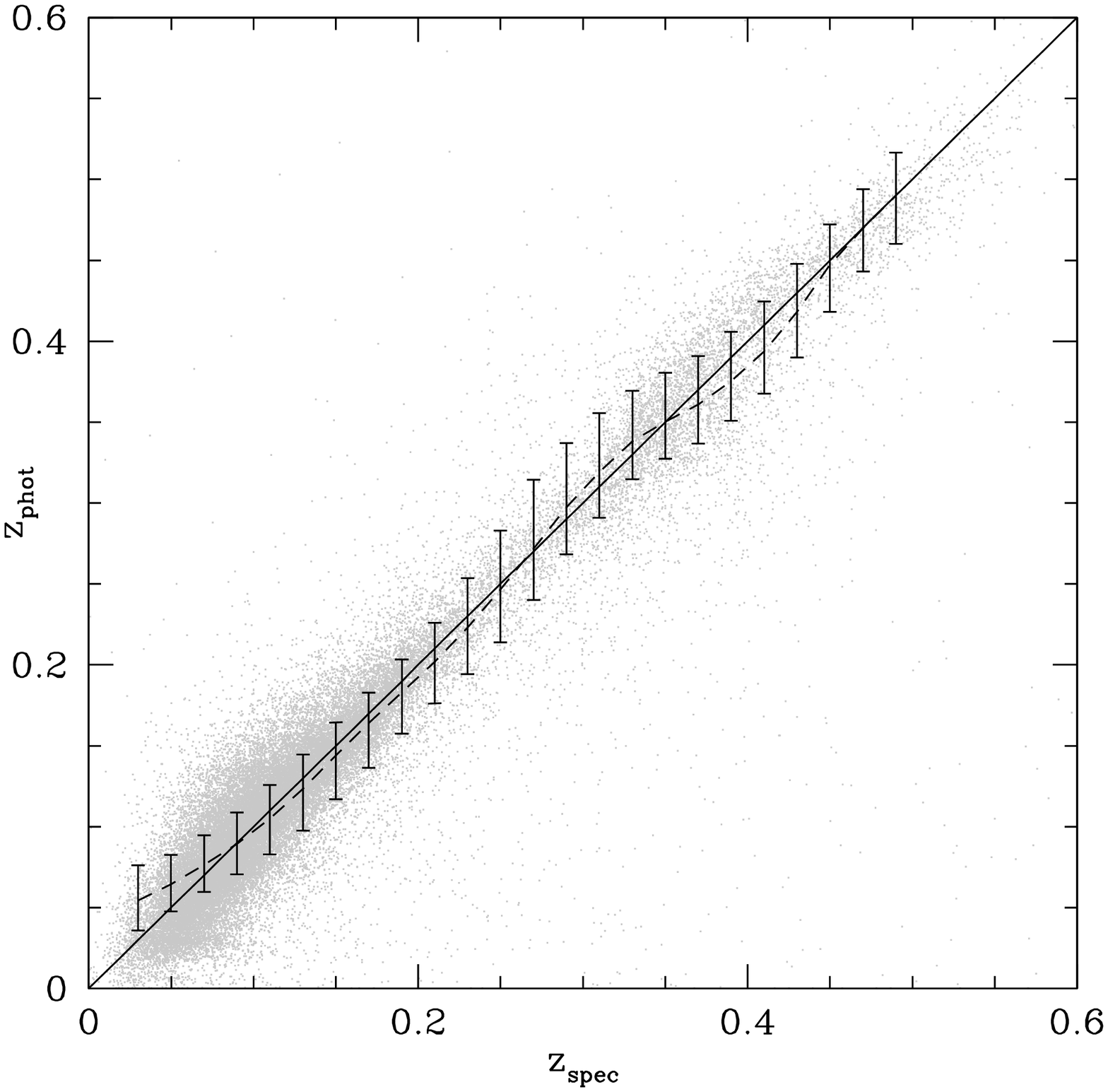,width=8.cm}}
\put(240,0){\psfig{file=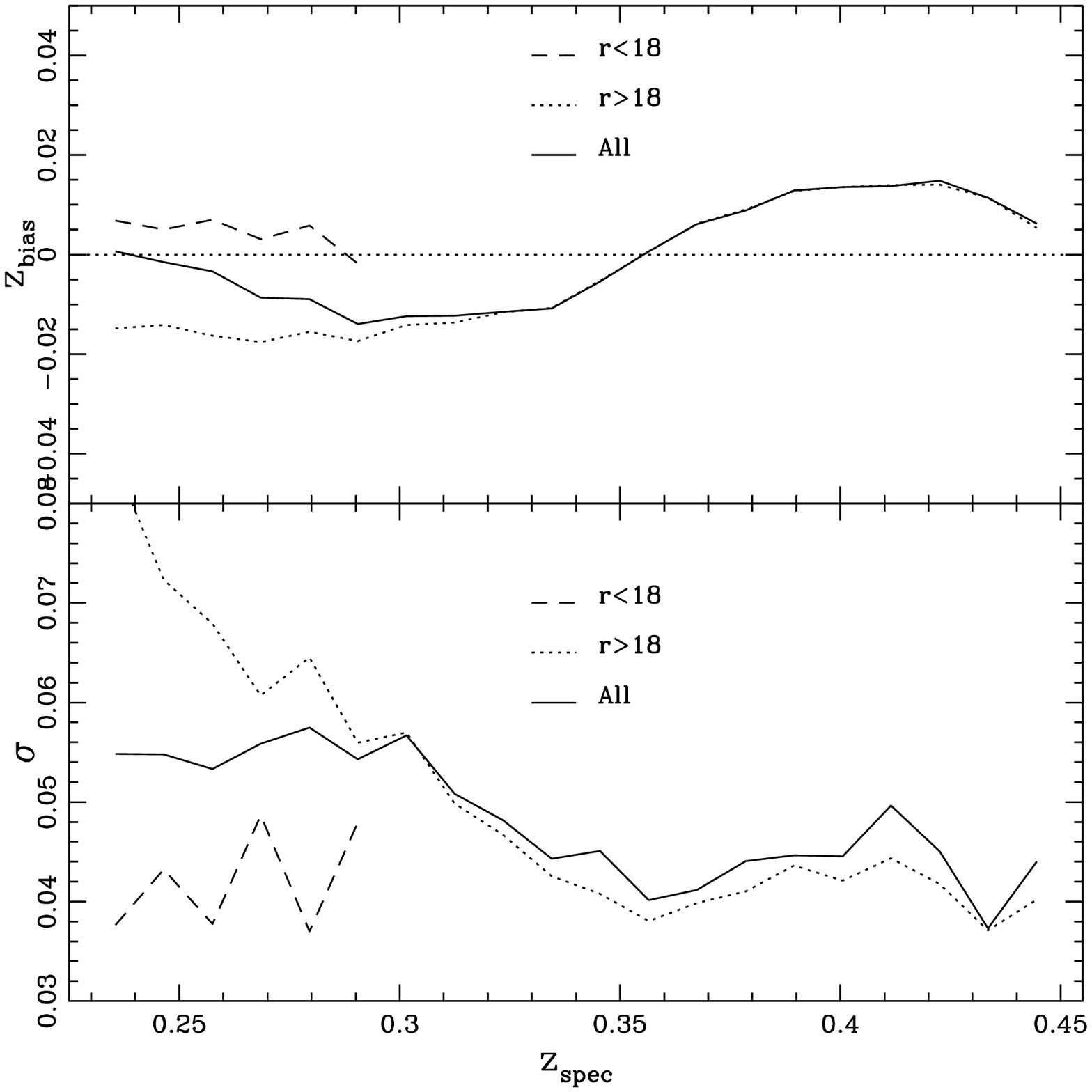,width=8.cm}}
\end{picture}
\caption{Left panel: $\zspec$ vs 
$\zphot$ relation for the SDSS-DR6 (dashed line);  errorbars correspond to the $10$ and $90$ percentiles. 
Values corresponding to individual galaxies
are shown as grey dots (shown for the $10\%$ of the sample). 
The
solid line shows the one-to-one relation.
The error bars correspond to 2$\sigma$ deviations.  Right panels: systematic (top right)
and stochastic (bottom right) errors in the photometric redshift estimates, for different apparent magnitude ranges
(see the figure key).}
\label{f1}
\end{figure*}

This paper is organised as follows. In Section 2 we describe the data used in our analysis, both observational
and from models of galaxy formation, Section 3 discusses the photometric redshifts available in the SDSS-DR6
and analyses their advantages and limitations.  Section 4 describes the method used to calculate projected galaxy densities,
Section 5 studies the evolution of galaxy colours as a function of their environment with redshift, and finally,
Section 6 summarises the results obtained in this work.

Throughout this paper, we adopt the cosmological model characterised by
the parameters $\Omega_m=0.25$, $\Omega_{\Lambda}=0.75$ and $H_0=70~h~
{\rm km~s^{-1}~Mpc^{-1}}$.

\section{The galaxy samples}
The samples of galaxies used in this work are drawn from the
catalogue of galaxies with photometric redshifts ($\zphot$) in the
SDSS-DR6 (\citealt{zp}, available in the SkyServer). 
These photometric redshifts are obtained using the Artificial Neural 
Network (ANN) technique to calculate $\zphot$ and the Nearest Neighbor 
Error (NNE) method to estimate $\zphot$ errors for $\sim 77$ million, 
$r<22$ objects classified as galaxies in the SDSS-DR6.  In the remainder 
of this paper, we restrict all our analysis to samples with $r<21.5$ since
this magnitude limit ensures a good photometric quality and a reliable 
star-galaxy separation (Stoughton et al., 2002, Scranton et al., 2002). 
We use the spectroscopic Main Galaxy Sample (MGS; \citealt{mgs}) to
test the reliability of photometric redshift estimates, and 
their ability in predicting projected local galaxy densities.
The MGS comprises all galaxies in the SDSS-DR6 brighter than a Petrosian 
magnitude $r_{\rm lim}=17.77$, with a median redshift $z \sim0.1$.

The SDSS-DR6 provides imaging data in five bands, \textit{ugriz},
and spectroscopy  over $\simeq \pi$ steradians in the northern Galactic 
cap, and $225$ square degrees in the south Galactic cap. 
The five broad bands \textit{u, g, r, i,} and \textit{z} 
\citep{fuku96,hogg01,smit02}, have effective wavelengths of $355$, 
$467$, $616$, $747$, and $892$ mm respectively. For details regarding 
the SDSS camera see \citet{gunn98}, for astrometric calibrations, 
see \citet{pier03}. SDSS-DR6 imaging data covers a contiguous region 
of the north Galactic cap. In this paper we use de-reddened model 
magnitudes K-corrected using V4.1 of the publicly available code described 
in Blanton \& Roweis (2006). We also repeat all the analysis presented in 
this work using alternative K-correction algorithms based on a grid of 
Bruzual \& Charlot (2003) models, and find that our results do not change 
in a significant way.

We will study the dependence of galaxy colour on environment for 
subsamples of different galaxy luminosities (samples L1 to L4, see
Table 1) and redshifts.  Given that galaxies at increasingly large 
distances subtend an ever smaller angle on the sky, it is expected that the 
size of the PSF of the image (due to seeing) will eventually become larger 
than the apparent galaxy size.  Such an effect would bias our sample 
against compact objects at large distances, in turn affecting our 
statistical measurements of the evolution of intrinsic galaxy properties.  
Therefore, we need to define our luminosity samples by taking into 
account both, the distance out to which we have a volume limited sample, 
and also the distance out to which the samples are free from selection 
effects due to the PSF.  The former is done by direct measurement on 
the Hubble diagram (absolute magnitude vs. redshift), while 
in order to obtain the latter we proceed as follows.  We use a low 
redshift sample composed by all galaxies in the range
$0.04<z<0.08$.  We divide this sample according to the luminosity limits 
that define our subsamples.  We measure $r_{10}$, 
the $10$ percentile of the distribution of physical galaxy radio,
and calculate the maximum redshift out to which $r_{10}$ subtends more 
than $1.5$ arcsec.  We take this limit as the typical seeing affecting 
the SDSS photometry (see for instance, Stoughton et al., 2002).  
This defines the maximum redshift ($z_{max}$) that ensures that compact 
galaxies in our samples are not confused with the PSF of the image.  
Table 1 also contains $r_{10}$ as well as this maximum redshift for 
each of our luminosity subsamples.  We will restrict the analysis to this 
maximum redshift from this point on; the table also shows the maximum 
redshift out to which the MGS (spectroscopic survey) can be used to 
construct volume limited samples, and as can be seen, $z_{max}$, the 
furthest we can reliably use the photometric catalogue, allows us a one
order of magnitude increase in sample size with respect to the MGS.

\begin{table*}
\begin{minipage}{175mm}
\caption{Definition of subsamples.  The second and third
columns show the maximum and minimum rest frame r-band absolute magnitudes of the sample, respectively.
The fourth column shows $r_{10}$, the first $10\%$ percentile of the galaxy size distribution at $z<0.08$.
The fifth, sixth and seventh columns are the redshift limits corresponding to volume-limited MGS, 
photometric $r<21.5$ sample, and the maximum redshift out to which galaxies can be identified without risk 
of confusion with the PSF using $r_{10}$, respectively. }
  \begin{center}\begin{tabular}{@{}ccccccc@{}}
  \hline
Sample & Minimum $M_r$ & Maximum $M_r$ & $r_{10}/$h$^{-1}$kpc & V. limited MGS $z$ & V. limited photometric 
cat.& $z_{max}$\\
 \hline
L1 & $-18.0$ & $-16.0$ & 1.8 & $0.02$ & $0.10$ & $0.08$ \\
L2 & $-19.5$ & $-18.0$ & 3.1 & $0.05$ & $0.23$ & $0.14$ \\
L3 & $-21.0$ & $-19.5$ & 4.7 & $0.09$ & $0.41$ & $0.23$ \\
L4 & $-21.5$ & $-21.0$ & 8.9  & $0.17$ & $0.74$ & $0.46$ \\
L5 & $-23.0$ & $-21.5$ & 11.7 & $0.21$ & $0.91$ & $0.63$ \\
\hline
\label{table:1}
\end{tabular}
\end{center}
\end{minipage}
\end{table*}

\subsection{Mock SDSS-DR6 catalogues}

We construct mock SDSS catalogues using the semi-analytic model (SAM) from Bower et al. (2006),
which is applied to the Millennium Simulation (Springel et al., 2005) outputs, to follow different
processes that shape the galaxy population as time progresses, corresponding to a $\Lambda$CDM cosmology.  The
Millennium simulation consists of $2160^3$ particles in a box of $500$h$^{-1}$Mpc a side, for
a particle mass resolution of $8.6\times 10^8$h$ ^{-1}$M$_{\odot}$.  The resulting galaxies
conform a complete sample down to a magnitude $M_r=-17$.

For each mock observer we apply the
same angular mask affecting the photometric SDSS-DR6 sample, and apply a magnitude limit cut of 
$r_{\rm lim}=21.5$ (observer-frame apparent magnitudes).  
We store all the observed properties of galaxies, such as redshift (which
includes the peculiar motion), angular position in the sky, and apparent magnitudes in several bands.  However,
the main advantage of the mock catalogue is that we also store several intrinsic properties such as the
luminosity in different bands.

We produce two separate mock catalogues down to $r_{lim}=21.5$.  
i) The first one uses the $z=0$ SAM output as the source of model galaxies out to the highest 
redshift, which is simply defined by the selection function of the
actual SDSS-DR6.  ii)  The second mock catalogue uses different outputs from the SAM
to take into account the evolution of the galaxy population out to $z=0.8$. 

In both cases, photometric redshifts are assigned following a Monte-Carlo procedure that replicates the
observed systematic and stochastic errors in the determination of photometric redshifts in the real SDSS-DR6.  We
acknowledge that this method only contains information from galaxies in the MGS, but since this includes a considerable
number of galaxies at redshifts $z>0.3$, we are able to assign photometric redshifts throughout the whole
redshift range.  We did not attempt to replicate the $\zphot$ algorithm applied to the real SDSS-DR6 due
to its complexity, and specially given the good observed behaviour of the error distribution of photometric
redshifts in the real data, which we review in detail in the following section.

\section {Photometric redshift estimates}

\begin{figure*}
\begin{picture}(450,250)
\put(0,0){\psfig{file=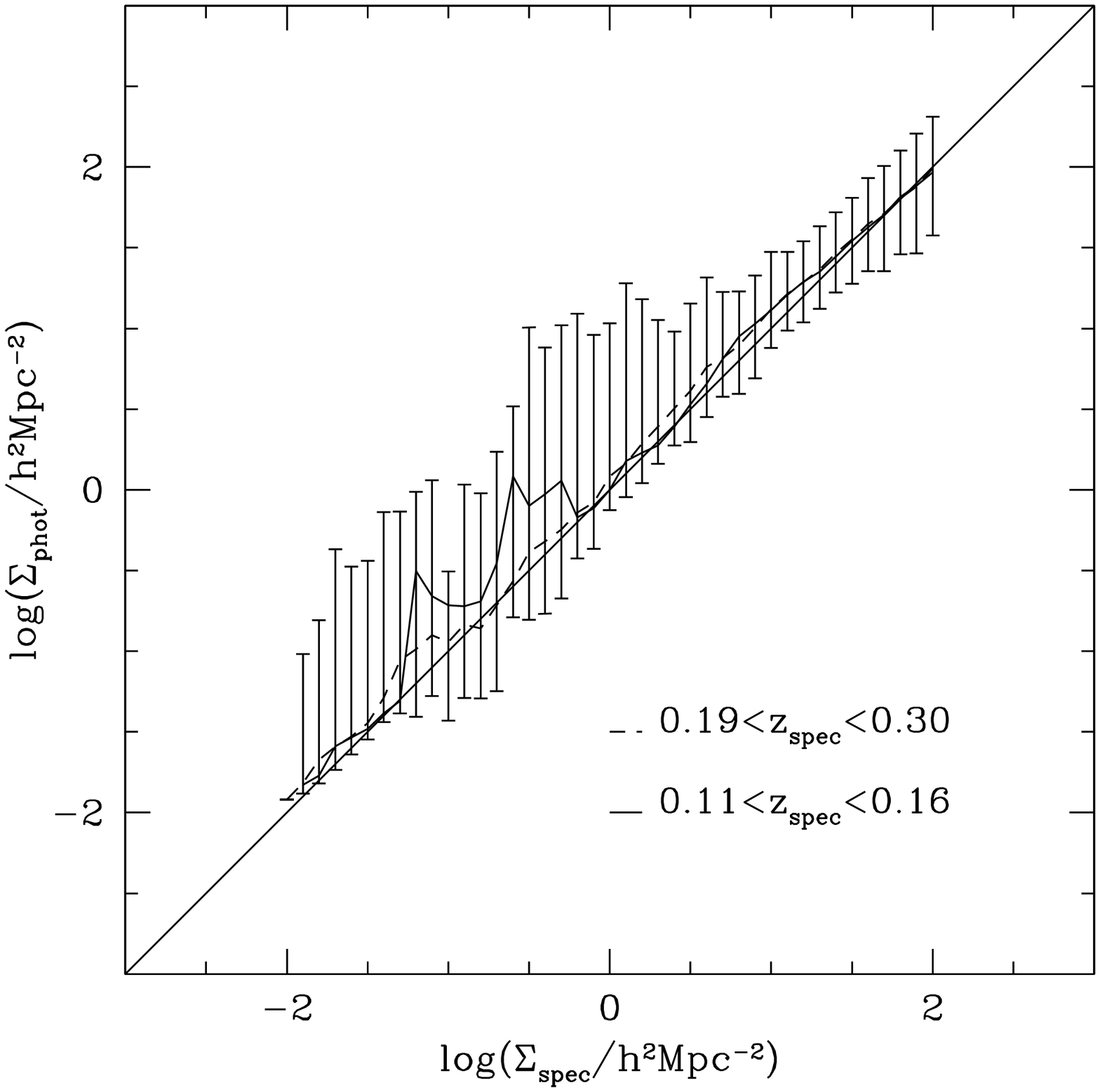,width=8.cm}}
\put(240,0){\psfig{file=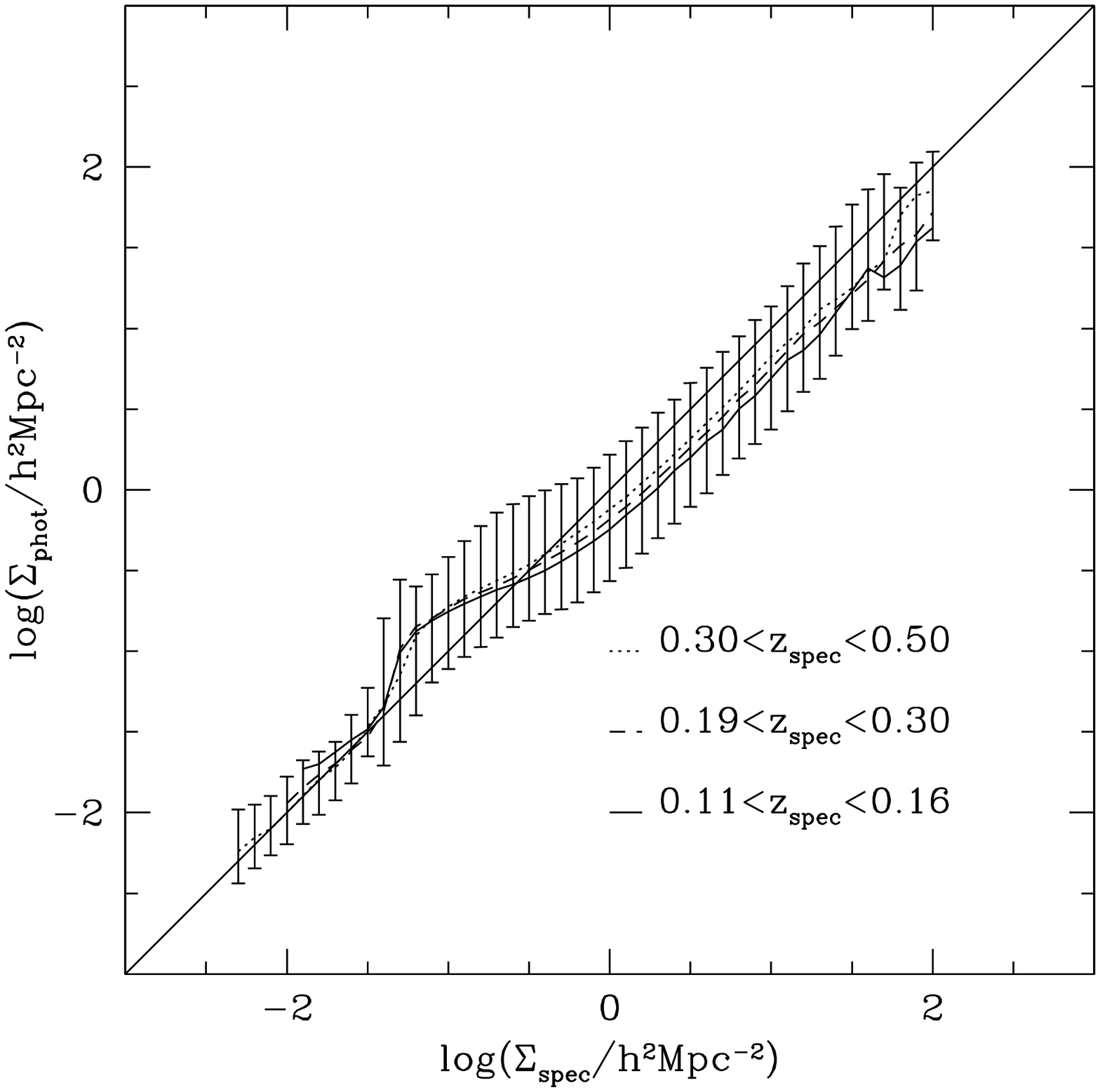,width=8.cm}}
\end{picture}
\caption{Correlation between spectroscopically ($\Sigma_{\rm spec}$) and 
photometrically ($\Sigma_{\rm phot}$) derived projected local galaxy densities, for different redshift ranges (shown in
the key). SDSS-DR6 and mock
results are shown in the left and right panels, respectively. The errorbars enclose $80\%$ of the 
distribution with the mean shown by thick solid lines.}
\label{f2}
\end{figure*}

Photometric redshift techniques use broadband photometry 
to perform an estimate of a galaxy redshift.
This technique can be used to infer large numbers of galaxy distances efficiently
including much fainter galaxies than
spectroscopic measurements which require using large telescopes. 
There exist different techniques to estimate photometric redshifts
which can be classified into two groups. The first set of techniques makes use of a small set of 
model galaxy spectra
derived from empirical- or model-based  spectral energy distributions. 
These methods reconstruct the observed galaxy colours
by finding the best combination of template spectra at different redshifts \citep{ben98,bolz00,csabai03}. 
The fact that these methods rely on a small number 
of template SEDs, is their main disadvantage, in particular for populations at 
higher redshifts.
On the other hand, there are techniques that need a large amount of prior redshift 
information (training set) called empirical methods \citep{con95,brun99}.
Their goal is to derive a parametrisation for the redshift as a 
function of the photometric parameters, inferred from the
training set.  The photometric parameters can be combinations of galaxy magnitudes in
different photometric bands, galaxy colours, and concentration indexes.

In this work we use the \citet{zp} $\zphot$ catalogue CC2, which uses colours and concentration indexes
as their training set.
The left panel of Fig. \ref{f1} shows the spectroscopic ($\zspec$) vs. $\zphot$
redshift for a $10\%$ of the objects in the SDSS-DR6 MGS.  The figure shows the one-to-one
relation as a solid line, and the median (dashed line), and $10$ and $9=$ percentiles (errorbars) 
of the scatter plot.  Interesting points
to notice are i) the lack of important concentrations of outliers, and ii) the small dispersion around the
one-to-one relation, indicating very precise estimates of redshifts.  This can be more clearly seen in
the right panel of this figure, where we show the systematic differences between $\zphot$ and $\zspec$, 
$z_{bias}$; the lower sub-panel shows the variance,$\sigma$, we notice that $\sigma \sim 0.05$ at 
$\zspec>0.3$. 

\begin{figure*}
\leavevmode \epsfysize=10.5cm \epsfbox{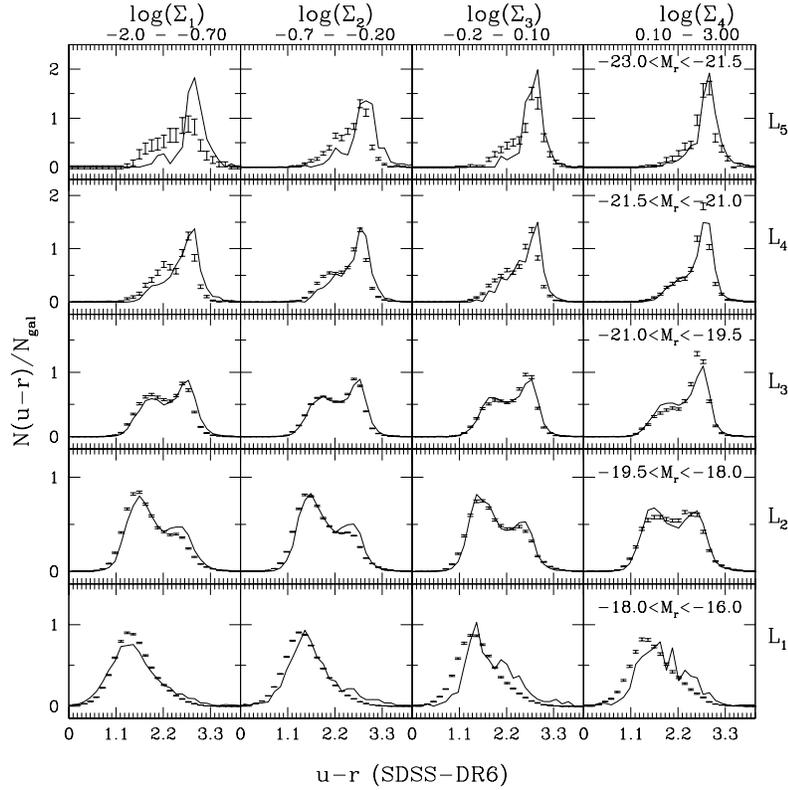}
\caption{ 
Distributions of $u-r$ colours in the SDSS-DR6 for different
ranges of projected density (increasing from left to right in each panel; the units, not shown
to improve clarity are $\left[\Sigma\right]=1/$h$^2$Mpc$^{-2}$) and
luminosities (increasing from bottom to top).  Small errorbars show the results
from photometric redshifts
and are computed assuming Poissonian statistics.  Solid lines show the results obtained using 
spectroscopic redshifts.  We only show galaxies 
with $z<0.08$.
}
\label{f3}
\end{figure*}

\section {Characterising Galaxy Environments}

We now address the accuracy of the use of photometric redshifts
to characterise galaxy environments. In order to do this
we use both photometric and spectroscopic redshifts in the SDSS-DR6 MGS to compute the 
projected galaxy densities using nearest neighbours in the plane of the sky.
This is done in both, real and mock catalogues using 
an extension of the $5-$th nearest neighbour local density estimator, $\Sigma_{5}$.  
In our case, since our samples are not volume-limited, we need to account for a varying 
average galaxy density as a function of redshift.  Therefore, we calculate the completeness 
of the survey at the redshift of a centre galaxy and use it to decide the most appropriate 
number of neighbours to calculate the density in such a way that these correspond to comparable 
volumes across the full range of redshifts.  In our case, the number of nearest neighbours, 
brighter than $M_r-5\log(h_{70})<-21$, is allowed to fluctuate
between $5$ and $10$; nearby galaxies will use larger numbers of neighbours than
galaxies at higher redshifts, where the completeness is lower.
The other important aspect of a projected density is that of selecting galaxies
within a given range of radial velocity difference ($\Delta V$). In the case of a spectroscopic survey,
one needs to take into account the velocity dispersion inside virialised structures rather than the
redshift measurement error (which is of the order of $100$km$/$s);
in photometric redshift surveys, on the other hand, the most important contribution to the smearing of
structure in the direction of the line of sight is the $\zphot$ error, which
can be larger than the expected virialised motions. 
We have tested the correlation between $\Sigma_{5}$ for spectroscopic and photometric redshift data 
in SDSS-DR6 ($z<0.3$) adopting a fixed $\Delta V =1,000$km$/s$ for $\zspec$ data and 
a varying $\Delta V$ in the range $1,000$-$3,000$km$/$s for $\zphot$,
taking into account 
the variation of the error in $\zphot$ with galaxy magnitude (taken from \citealt{zp}).
The results of this study
show that the best correlation occurs for a value of $\Delta V$ corresponding to 
$\simeq 30\%$ of the error in the photometric redshift estimate.
The maximum velocity difference allowed in our analysis is $4,000$km$/$s.
These improvements to the projected density estimator allow us to significantly improve the 
accuracy of the density estimates using photometry only data.

\begin{figure*}
\begin{picture}(450,240)
\put(0,0){\psfig{file=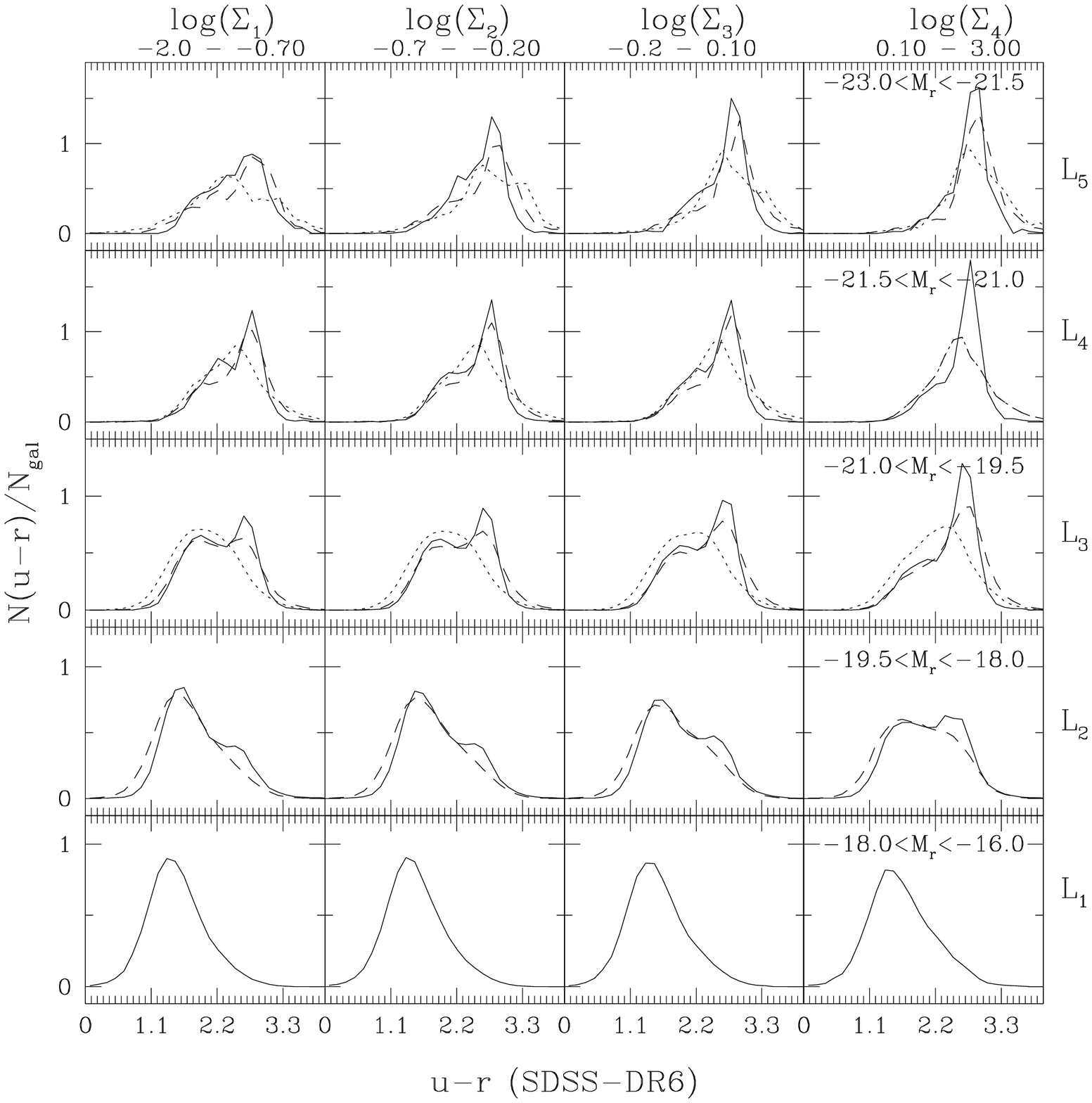,width=8.cm}}
\put(240,0){\psfig{file=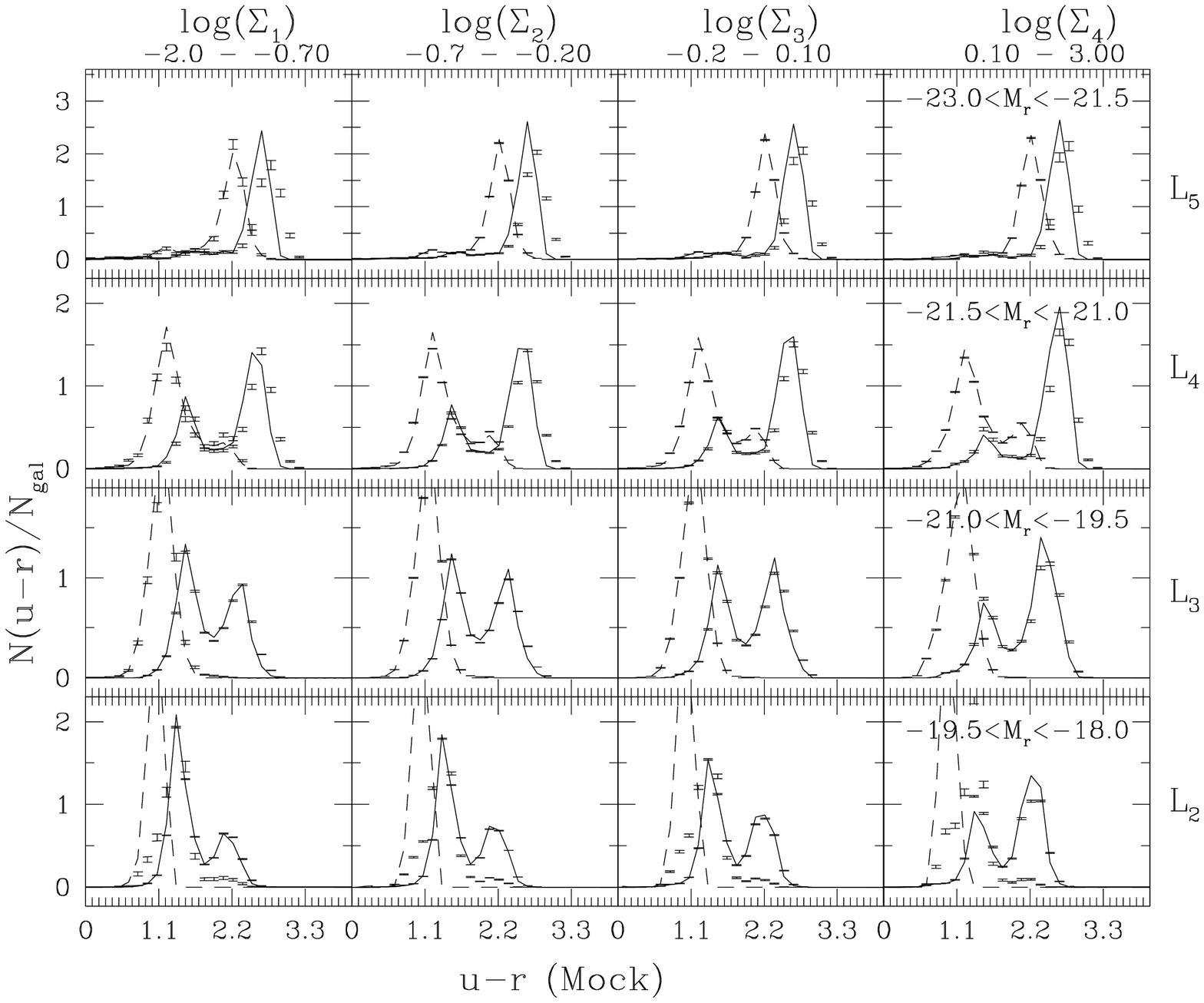,width=8.cm}}
\end{picture}
\caption{Colour distributions for SDSS-DR6, and evolving mock catalogues (left and right, respectively)
as a function of luminosity (bottom to top, see the key in the right-most subpanels) and local
density (left to right, units are as in Figure \ref{f3}), for different median redshifts $z=0.062,0.124,0.19$
(solid, dashed and dotted lines, respectively, left panel only) and $z=0.08$ and $0.5$ (solid and dashed lines,
respectively, right panel only).
}
\label{f4}
\end{figure*}

Figure \ref{f2} shows the relation between projected densities calculated using
spectroscopic (x-axis, $\Sigma_{\rm spec}$) and photometric (y-axis, $\Sigma_{\rm phot}$) 
redshift estimates in the SDSS-DR6 (left) and the no-evolution mock (right) catalogues, for different
redshift ranges (solid and dashed for low and higher redshifts, respectively; additionally, the mock
catalogue results for even higher redshifts are shown in dotted lines).
Errorbars enclose $80\%$ of the distribution and is only shown for the high redshift densities
(the dispersion in the relation does not show important changes with redshift), 
and the solid line shows the one-to-one relation.  As it can be seen, the SDSS and mock
catalogues show a very good agreement between spectroscopic and photometric projected
densities, although there is a slight tendency for SDSS-DR6 photometrically derived densities to 
show a high density tail.  Notice that the errors in the projected densities shown for the mock
correspond to a much higher redshift than the SDSS-DR6; this is to check that the estimates
of density are also accurate at redshifts higher than those accessible using the SDSS-DR6 MGS. 
Another effect that should be taken into account is the SDSS-DR6 minimun 
fiber separation ($55"$) which could affect the results from the spectroscopic data. 
We use the mock catalogue to study this effect on the deepest spectroscopic sample we use in this 
work, $z \sim 0.3$, and find that $\Sigma_{spec}$ can be underestimated by only an average of a $1\%$.


\begin{figure*}
\begin{picture}(450,250)
\put(-20,0){\psfig{file=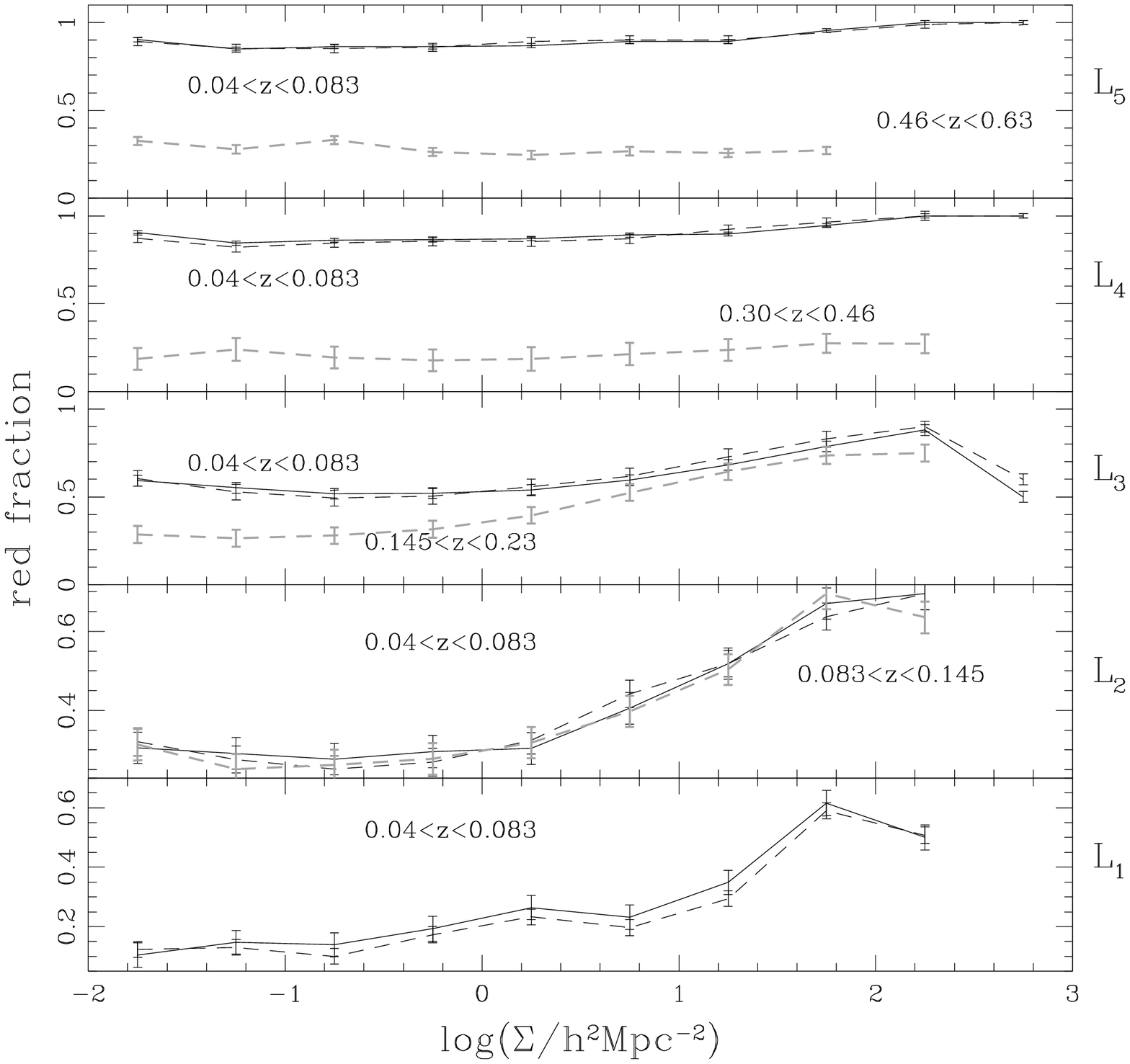,width=8.5cm}}
\put(230,0){\psfig{file=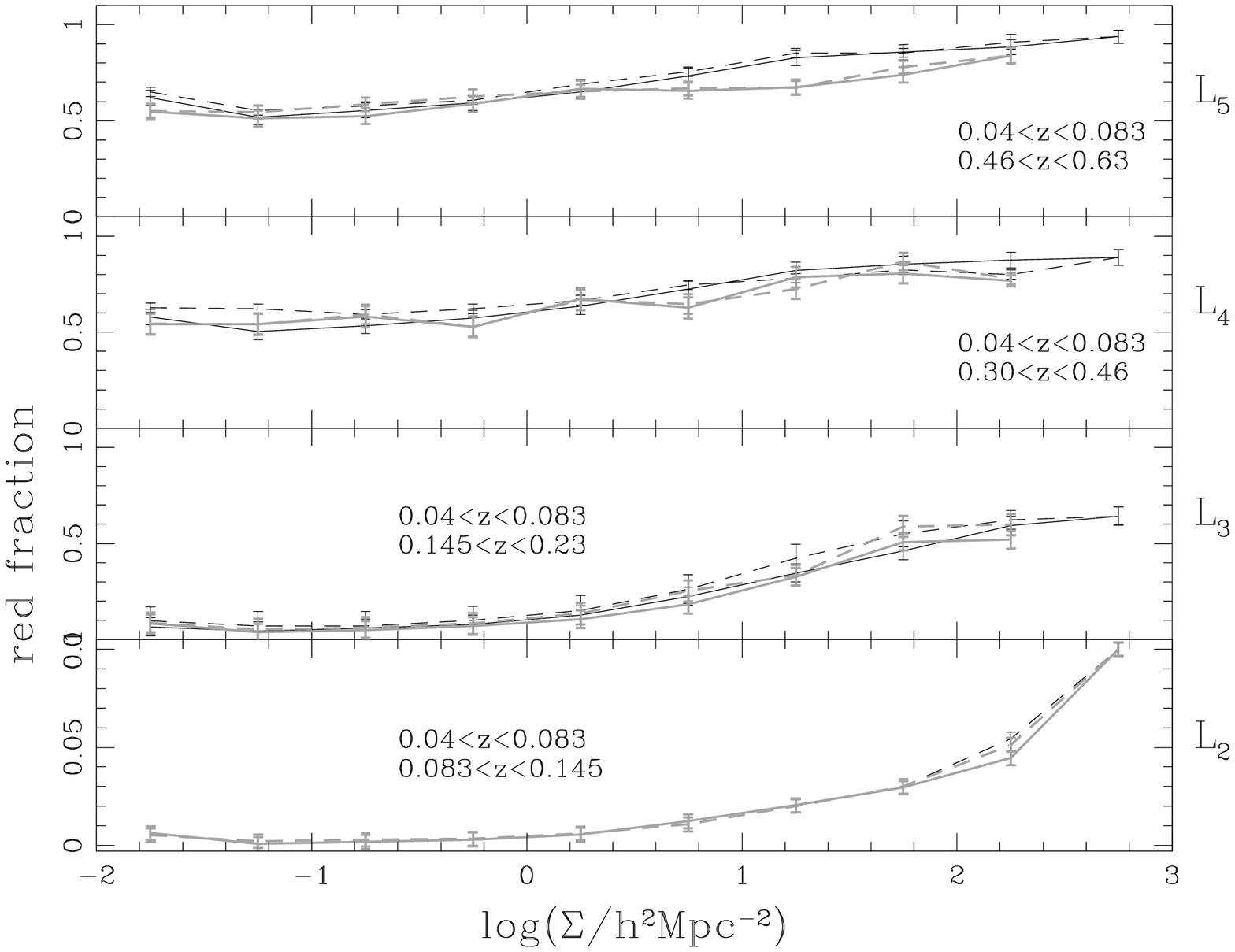,width=8.5cm}}
\end{picture}
\caption{ 
Red galaxy fractions in the SDSS-DR6 and mock catalogues (left and right panels, respectively) 
for different luminosities (increasing from bottom to top sub-panels), as a function of projected
density, $\Sigma$, and in different redshift bins (see the figure key). Dashed lines correspond to the red 
galaxy fraction on the photometric sample for high and low redshift (grey and black lines respetively) 
and solid lines correspond to the results from the spectroscopic sample.
}
\label{f5}
\end{figure*}

\section{Evolution of environment dependent galaxy properties}

We now explore the effects of environment on galaxy colours at different redshifts.  We start
by comparing the variation of the distributions of colours
as a function of spectroscopic and photometric densities.  This comparison can only be
performed at low redshifts in the SDSS-DR6.  
Figure \ref{f3} shows the colour distributions
for the SDSS-DR6, for different luminosity and density bins (different subpanels, 
see the figure key). Solid lines show the distributions obtained using spectroscopic 
projected densities; small errorbars computed using Poisson statistics,
describe the distributions measured using photometric 
densities.  We have also computed an equivalent colour distribution for the mock catalogues
finding that spectroscopic and photometric density estimates are almost 
indistinguishable. The slight differences in the SDSS-DR6 do not affect the fraction 
of red galaxies inferred from these diagrams, as it will be shown later in this section. 
This confirms that photometrically derived densities can be suitable to study the dependence 
of galaxy colours on environment.

The galaxy population in the low redshift SDSS-DR6 data shows the previously reported behaviour
\citep{balogh04,baldry04}, where
low luminosity galaxies show a transition from a blue, unimodal  color distribution to a slightly
bimodal distribution including a population of red galaxies as the local density increases.  Intermediate
luminosities show a clear transition from a blue dominated population to a mostly red population, and
high luminosity galaxies show an increasing deficit of blue galaxies.  The mock catalogue, on the other
hand, shows narrower colour distributions with qualitatively similar behaviours to the observed data. 
As it has been reported in previous works (cf.  Weinmann et al., 2006), the colours in semi-analytic
models still fail to reproduce observed values with high accuracy.  We do not show colours for low
luminosity semi-analytic galaxies due to the completeness limit of the simulation.

As the redshift increases, we find that SDSS-DR6 galaxies show different behaviours depending on their
luminosity.  The left panel of Figure \ref{f4} shows, with different line types, the colour distributions 
corresponding to
different redshift subsamples up to a maximum median redshift of $z=0.19$; 
local densities in this figure are obtained using
photometric information only. As can be seen, for the  high luminosity galaxies (L5), 
there is no statistically significant trend with local density, regardless of the redshift of the 
sample; there is a clear trend towards bluer L5 galaxies at higher redshifts, where the red galaxy fraction drops
by a $(60\pm0.05)\%$ between median $z=0.052$ and median $z=0.55$.  Fainter galaxies (L2) show a more noticeable 
shift towards bluer colours, which at low redshifts show the transition from a 
blue-dominated population to a main population
of red galaxies, a $(38\pm0.07)\%$ effect. 
This seems to be the case also for more intermediate luminosities (L3), which show a similar trend with
local density (a $(30\pm10)\%$ effect), which becomes clearer at higher redshifts (a $(50\pm10)\%$ increase
in the red fraction for higher local densities).
These results are compatible with those presented by Cucciati et al. (2006), De Lucia et al., (2007) and
\citet{roser}, who also find a rapid migration to bluer colours as the redshift increases.

The right panel of Figure \ref{f4} also shows in different line types the colour distributions for two different
redshifts, $0.04<z<0.12$ and $0.45<z<0.55$) 
obtained from the evolving mock catalogue using spectroscopically defined densities (errorbars 
were obtained using photometric density estimates).  The main purpose of this panel is to show that
colour distributions of galaxies selected according to their photometric estimates of local density, are compatible
with those obtained using spectroscopic densities, even at high redshifts.  On the other hand, it is noticeable
that the evolution of colours in the mock catalogue is qualitatively similar to that in the SDSS in the sense that
low luminosity galaxies at high redshifts show a lack of a red galaxy population.  The most luminous galaxies
in the model become much bluer than observed in the SDSS-DR6.
We now turn to make quantitative estimates of the changing colours of the general galaxy population.

A more direct way to characterise the distribution of galaxy colours is via estimates of the red galaxy
fraction.  This is defined as the ratio between the number of galaxies redder than $u-r=2.2$ and the
total number of galaxies.  Figure \ref{f5} shows the dependence of this fraction on local density for different
galaxy luminosities (increasing luminosity from bottom to top subpanels) and different redshifts (different
line types, redshift ranges indicated in the key), for the SDSS data and the no-evolution mock catalogue (left
and right panels, respectively).  In all subpanels, we show two redshift ranges and spectroscopically 
and photometrically derived densities. Notice that in the SDSS-DR6 case,
we only have spectroscopic data for the low redshift range.
As can be seen, the results of the no-evolution mock using either spectroscopic or photometric 
density estimates show a very good agreement with the underlying lack of evolution, indicating 
that there are virtually no differences in the red fractions using these two density estimators.
In the case of the SDSS-DR6, there is a significant drop in the
red galaxy fraction as the redshift increases.  For the highest luminosity galaxies 
(labeled L5), the fraction plummets from a value of $\simeq 0.9-1$ to $\simeq 0.3$, 
remaining constant as the local density increases.  Intermediate luminosities (labeled L3), 
show that the red fraction drops by a $50\%$ between median redshifts of $z=0.062$ and $z=0.19$, 
for low surface densities, $\Sigma/$h$^2$Mpc$^{-2}<1$; higher densities show a less significant
change with redshift.

\begin{figure}
\leavevmode \epsfysize=8.5cm \epsfbox{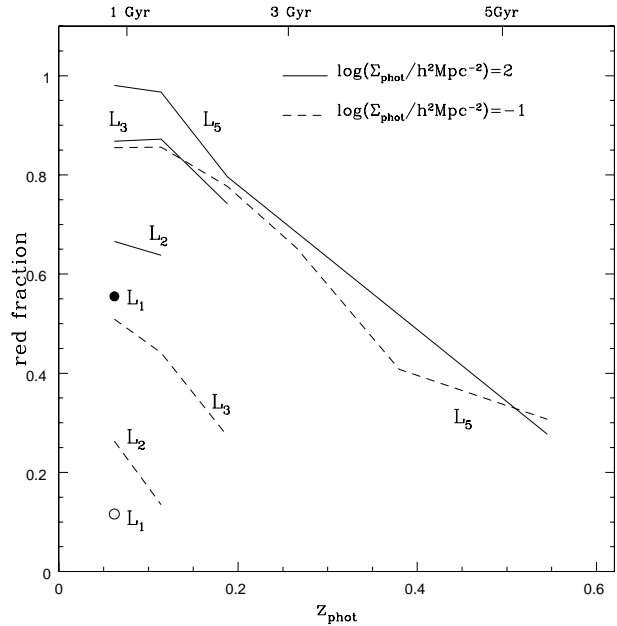}
\caption{ 
Evolution of the red galaxy fraction in the SDSS-DR6, for two different projected
densities (see the figure key).
}
\label{f6}
\end{figure}

\section{Conclusions}

We have studied the evolution of the distribution of galaxy colours in the SDSS-DR6  at intermediate 
redshifts $0.04<z<0.63$, and its dependence on projected local density, as 
inferred from nearby neighbours in projection, using photometric redshift information.

A first step consisted in constructing mock catalogues with photometric redshifts and uncertainties derived
from the differences between $\zphot$ and $\zspec$ in the SDSS-DR6 MGS, in two versions, one with an evolving
galaxy population and another one with no evolution.  These mocks are used for two main purposes, i) to determine
whether the statistical analyses are affected by systematic errors, and to assess stochastic errors
to improve the determination of projected densities for
the analysis on real data, and ii) to compare the evolution predicted by the galaxy formation model
to the observed evolution. 

Using the SDSS-DR6 MGS, we are able to tune our density measurement technique as well as to
determine that the estimates of projected densities using photometric
redshift information are in agreement with the results using spectroscopic data, out to $z\simeq0.3$.  We confirm
this result in the mock catalogues (both with and without evolution), and extend this test out to much higher
redshifts, $z=0.63$, the maximum redshift out to which we can analyse the SDSS-DR6 photometric catalogue
avoiding possible confusion problems between galaxies and the PSF.
We also test whether the colour distributions of galaxies selected in bins of projected density are comparable
when using photometric and spectroscopic distances to select close neighbours for the determination of projected
densities.  We find that the distributions are compatible both in the real and mock data.  In this case, the variations
in the colour distributions seen in the model are narrower than in the real data, but follow qualitatively 
similar trends with
density and luminosity.  Low luminosity galaxies show mostly a blue population, with signatures of a red population
appearing only at the highest density bin; high luminosity galaxies show mostly a red population, with some blue 
galaxies only at low densities.  Intermediate luminosities show a nice transition from a bimodal colour distribution
at low densities to an almost unimodal red distribution at high densities.  These results are in excellent
agreement with previous studies (e.g. Balogh et al., 2004).

The evolution of the colour distributions with redshift shows that low luminosity galaxies tend to loose the red
population almost completely, while the blue population shifts to even bluer colours; high luminosity galaxies, 
on the other hand, show a constant shift to bluer colours; this is more easily seen in our study
of red galaxy fractions for different redshift intervals.  One important characteristic of this evolution
is that  the dependence of the red fraction on local density in high luminosity
galaxies, $M_r-5\log(h_{70})<-20.5$,
retains its overall shape and only changes by a constant
shift, 
which for $-23<M_r-5\log(h_{70})<-21.5$, corresponds to a
$\simeq 60\%$ increase from median $z=0.545$ to $z=0.05$.
Intermediate luminosities, $-21<M_r-5\log(h_{70})<-19.5$, show a $50\%$
increase from a median $z=0.19$ to median $z=0.05$ in the red fraction 
in relatively low projected densities; galaxies in high density environments show a milder 
change in the red fraction in this redshift range.
This evolution can be more easily seen as a function of redshift for fixed values of local density, 
as is shown in Figure \ref{f6}, where solid lines and filled symbols correspond to high 
densities, and dashed lines and open symbols to very low local densities.  
As can be seen, the slope in the red fraction evolution is approximately constant 
implying that in the local universe, the blue fraction decreases at a rate of $15\%/$Gyr.  
There is an indication that faint and bright galaxies follow a similar trend, but due to the magnitude 
limit of the SDSS-DR6, the data on the evolution of the fainter subsamples reaches lower redshifts, 
increasing the uncertainty in the estimate of the slope.

These results are consistent with previous works (Cucciati et al., 2006, De Lucia et al., 2007, 
\citealt{martinez}, Ienna \& Pell\'o, 2006),
where the population of high luminosity galaxies becomes red at higher redshifts than their low luminosity counterparts.
This phenomenon has been termed ``downsizing" in the literature
(e.g. \citealt{marconi04}, \citealt{shankar04}),               
in relation to the star formation activity shifting from larger objects
towards smaller galaxies, in apparent contradiction to the hierarchical model of galaxy formation.  Several
recent works have shown that this behaviour is in fact also natural in a hierarchical Universe
(Lagos, Cora \& Padilla, 2008, Bower et al., 2006, Croton et al., 2006); however, it still remains
a challenge for galaxy formation models to reproduce the actual distribution and evolution of galaxy colours
which at present is achieved only qualitatively at best,
for which this work provides the most accurate measurements carried out to date.
\section{Acknowledgments}
We thank referee for very helpful comments.
NP was supported by Fondecyt grant No. 1071006.  The
authors benefited from a visit of DGL to Santiago de Chile supported by
Fondecyt grant No. 7070045.  This work was supported in part by the
Centro de AstrofÌsica FONDAP, Consejo Nacional de 
Investigaciones Cient\'ificas y T\'ecnicas de la Rep\'ublica Argentina 
(CONICET), Secretar\'\i a de Ciencia y Tecnolog\'\i a de la Universidad 
dad de C\'ordoba.
Funding for the SDSS and SDSS-II has been provided by the Alfred P. Sloan Foundation, 
the Participating Institutions, the National Science Foundation, the U.S. 
Department of Energy, the National Aeronautics and Space Administration, 
the Japanese Monbukagakusho, the Max Planck Society, and the Higher Education 
Funding Council for England. The SDSS Web Site is http://www.sdss.org/. 
The SDSS is managed by the Astrophysical Research Consortium for 
the Participating Institutions. The Participating Institutions are 
the American Museum of Natural History, Astrophysical Institute 
Potsdam, University of Basel, University of Cambridge, Case 
Western Reserve University, University of Chicago, Drexel 
University, Fermilab, the Institute for Advanced Study, the Japan 
Participation Group, Johns Hopkins University, the Joint Institute 
for Nuclear Astrophysics, the Kavli Institute for Particle 
Astrophysics and Cosmology, the Korean Scientist Group, the 
Chinese Academy of Sciences (LAMOST), Los Alamos National 
Laboratory, the Max-Planck-Institute for Astronomy (MPIA), the 
Max-Planck-Institute for Astrophysics (MPA), New Mexico State 
University, Ohio State University, University of Pittsburgh, 
University of Portsmouth, Princeton University, the United States 
Naval Observatory, and the University of Washington.


\label{lastpage}

\end{document}